# High-Q Two-dimensional Photonic Crystal Nanocavity on Glass with an Upper Glass Thin Film

Ryusei Kawata[1*], Akinari Fujita[1], Natthajuks Pholsen[2], Satoshi Iwamoto[2] and Yasutomo Ota[1†]

[1] *Department of Applied Physics and Physico-Informatics, Keio University, Yokohama, Kanagawa 223-8522, Japan*
[2] *Research Center for Advanced Science and Technology, The University of Tokyo, Meguro, Tokyo 153-8904, Japan*
**r.kawata@keio.jp*, †*ota@appi.keio.ac.jp*

**We numerically analyze two-dimensional photonic crystal (PhC) nanocavities on glass with a thin glass film on top of the structure. We investigated a multi-step heterostructure GaAs PhC nanocavity located on glass. We found that covering the structure even with a very-thin glass film efficiently suppresses unwanted polarization mode conversion occurring due to the asymmetric refractive index environment around the PhC. We also uncovered that the glass-covered structure can exhibit a higher $Q$ factor than that observed in the structure symmetrically cladded with thick glass. We point out that the mode mismatch between the PhC nanocavity and modes in the upper glass film largely contributed to the observed $Q$-factor enhancement. These observations were further analyzed through the comparison among different types of on-glass PhC nanocavites covered with thin glass films. We also discuss that the in-plane structure of the upper glass film is important for additionally enhancing $Q$ factor of the nanocavity.**

Photonic crystal (PhC) nanocavities realize strong light confinement both in time and space and are a fascinating platform for the development of functional and compact photonic devices[1–3]. High $Q$-factor optical resonances with small mode volumes ($V$s) supported in PhC nanocavities[4–7] significantly enhance light matter interactions. The photonic bandgap effect can suppress undesired spontaneous emission from emitters embedded in the structure[8]. These properties are advantageous for the development of low threshold nanolasers[9–11], compact nonlinear optical devices[12–15], and efficient quantum light sources[16,17]. To integrate these nanophotonic devices on chip, an important development is the realization of high $Q$-factor cavity resonances on glass. Efficient evanescent optical coupling from the cavity to a glass-clad waveguide is possible only when the cavity exhibits a high unloaded $Q$ factor on glass. Moreover, on-glass structures are important for achieving mechanically robust PhC nanocavities.

One of the major approaches for realizing high-$Q$ PhC nanocavities on glass is to use one dimensional (1D) PhC nanobeam structures[18–20]. $Q$ factors over hundreds of thousands with $V$s close to the diffraction limit have been demonstrated at telecommunication wavelengths in the structures formed in silicon-on-insulator substrates[19]. Even using compound semiconductors, it is possible to realize high $Q$-factor 1D cavities on glass. Such III-V semiconductor-based 1D PhC nanocavities have been demonstrated to be efficiently coupled to underlying waveguides cladded with glass[21–24]. While the 1D PhC nanocavities show success in terms of on-chip integration, they only show limited photonic bandgap effect, which could hamper the realization of highly-efficient light sources. Moreover, the 1D structures are in general inferior to heat dissipation than other higher dimensional structures, which may limit their use in certain applications.

The above drawbacks listed for the 1D PhC structures could be resolved by the use of 2D PhC nanocavities. They support stronger photonic bandgap (PBG) effect and enables the efficient funneling of spontaneous emission from emitters into the target cavity modes. The 2D structure could facilitate heat dissipation from the cavity region. However, it is widely known that the $Q$ factor of a 2D PhC nanocavity tends to be significantly deteriorated when being placed on glass, due to the mode conversion between transverse-electric (TE)-like and transvers-magnetic (TM)-like modes and the resulting optical loss from the TE polarized cavity modes. The TE-TM mode conversion occurs due to the asymmetric refractive index distribution around the cavity[25,26]. A simple idea to achieve high $Q$-factor 2D PhC nanocavities is to symmetrically clad the structure by spin-on-glass (SOG) materials. Such a cavity design based on a multistep heterostructure PhC resulted in an ultra-high $Q$ factor of $8\times10^7$ with a small $V$ of $1.6(\lambda/n)^3$[27]. However, the design requires precise control of the refractive index of the SOG to achieve a high $Q$ factor. Another technique to realize high $Q$ factor on glass is the gentle spatial confinement of cavity field, which effectively suppresses the TE-TM mode conversion. Even without an upper clad, a PhC nanocavity based on waveguide width modulation showed a high $Q$ factor of $1.2\times10^7$[28]. However, this approach sacrifices spatial light confinement and the resulting $V$s tend to be relatively large.

In this study, we investigated high-$Q$ 2D PhC multistep heterostructure nanocavities on glass by placing a thin glass film on top of the structure. We found that the additional thin glass effectively suppresses the TE-TM mode conversion and thus largely improves cavity $Q$ factor. Interestingly, the thin-glass-covered structure exhibits a 1.7-times higher $Q$ factor than that observed in the structure with thick symmetric glass clad. We analyzed the mechanism of the increased $Q$ factor and examined the glass-film-loaded structures with different types of 2D PhC nanocavities. We also found that shaping the loaded glass in the in-plane direction could additionally improve $Q$ factor.

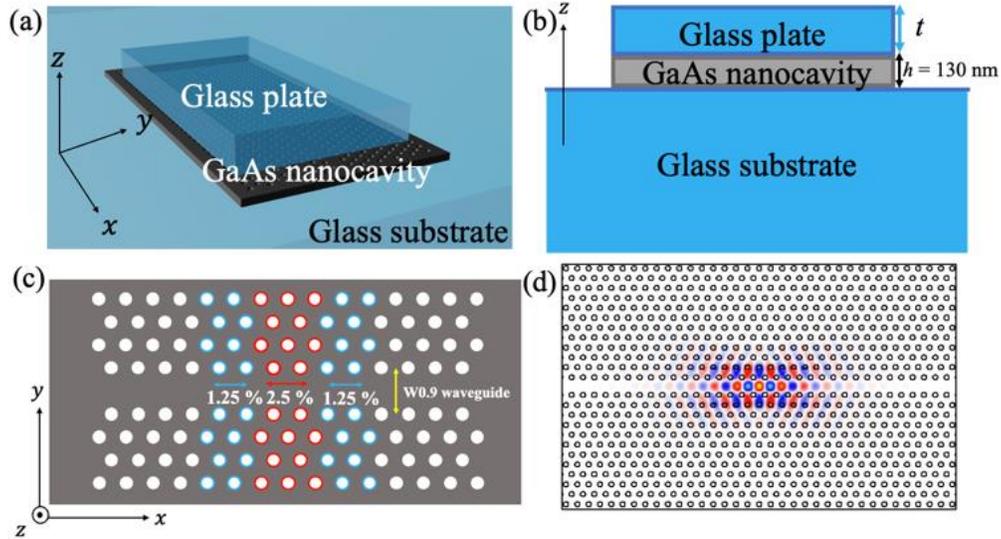

Fig.1. (a) Bird's-eye and (b)cross-sectional view of the schematic of the investigated structure. A 2D PhC nanocavity is placed on a glass substrate, and a glass thin film is loaded on top of it. (c) Top view of the heterostructure 2D PhC nanocavity and (d) the electric field distribution ($E_y$) of the cavity mode. Lattice constant modulations in $x$-direction are applied for a W0.9 waveguide in the PhC. The lattice constants are widened in the $x$-direction by 2.5% for the airholes colored red and 1.25% for blue.

Figure 1(a) shows a schematic of the investigated multistep heterostructure 2D PhC nanocavity on glass. The GaAs 2D PhC slab with a fixed thickness $h$ of 130 nm is placed on a thick glass substrate and further covered with a glass thin film with a variable thickness of $t$. A cross section of the investigated structure is shown in Fig. 1(b). The upper and bottom glass layers are assumed to share the same refractive index ($n_{Glass}$ = 1.45). Figure 1(c) shows the design of the multistep heterostructure PhC cavity[5]. The lattice constant $a$ is 240 nm and the radius of the air holes $r$ is 0.22$a$. The cavity is formed in a W0.9 waveguide in the GaAs PhC ($n_{GaAs}$ = 3.45) by introducing a multistep heterostructure by elongating the lattice constant by 2.5% and 1.25% in *x* direction (see the holes colored in red and blue). The cavity shows a high $Q$ factor of $4.9\times10^7$ and a small $V$ of $1.7(\lambda/n)^3$ at the resonance wavelength of $\lambda = 918$ nm when being fully cladded by air. The tight light confinement can be confirmed in a computed field distribution of the investigated cavity mode shown in Fig. 1(d). We numerically analyzed the cavity by the 3D finite-difference time-domain (FDTD) method. The computational domain is enclosed by perfectly matched layers (PMLs). The distance between the GaAs surface and the upper PML is set to be 1.5 µm, which corresponds to the maximum value of $t$.

First, we investigate the behavior of $Q$ factor of the PhC nanocavity covered with the glass thin film. Here, the PhC slab and the glass film laterally extend to the edge of the computation domain. Figure 2(a) summarizes computed $Q$ factors as a function of glass thickness $t$. When $t = 0$, which corresponds to the nanocavity nakedly placed on glass, we observed a very low $Q$ factor of about $10^4$, which is roughly three orders of magnitude lower that the $Q$ of the air-clad structure. The significant reduction of $Q$ factor arises from the TE-TM mode conversion due to the asymmetric refractive index environment. The TM polarized light converted from the TE polarized intra-cavity light does not feel any PBG effect and thus quickly leaks out from the cavity. In this case, the cavity loss rate is governed by the rate of the mode conversion. When $t > 1$ µm, which corresponds to the nanocavity sandwiched by thick glass bulks, we observed a saturation of $Q$ factor to a significantly improved value of $2.5 \times 10^5$. In these cases, the upper glass layer is much thicker than the wavelength of light in glass ($\lambda/n_{Glass}$) and the TE-TM mode conversion is predominantly inhibited owing to the recovered symmetry. Unexpectedly, the $Q$ factor of the nanocavity takes its maximum value of $4.2\times10^5$ when $t = 260$ nm, which is much thinner than ($\lambda/n_{Glass}$). The maximum $Q$ factor is 1.7 times higher than that of $t > 1$ µm.

To understand the observed improvement of $Q$ factor, we analyzed the loss mechanism of the cavity mode by monitoring the power distribution of light leakage from the cavity. We located three power monitors around the computational domain edges as shown in the inset of Fig. 2(b). The side monitor with a height of $2h$ is placed at the PhC slab center and detects light leaking through the PhC slab and its vicinity. Figure 2(b) summarizes *t*-dependences of light leakage power measured at the top, side and bottom detectors, together with total loss rates of the cavity mode. For small *t*s less than 200 nm, the optical loss is dominated by the lateral light leakage (blue curve) due to the TE-TM mode conversion. This is evidenced by the cross-sectional field distribution plotted in the panel of $t = 0$ µm in Fig. 2(c). Waveguiding of the TM polarized light in the GaAs layer is clearly seen in the plot. We observed exponential decrease of the lateral leakage when increasing $t$ (see the blue curve in Fig.2 (b)). This tendency can be well explained by a coupled mode theory assuming that the TE cavity mode mainly interacts with the guided TM waves, which exponentially decay in the air region[29]. In contrast, the light leakage into the lower direction is less dependent on *t*, which can also be confirmed in the field distributions plotted in Fig. 2(c).

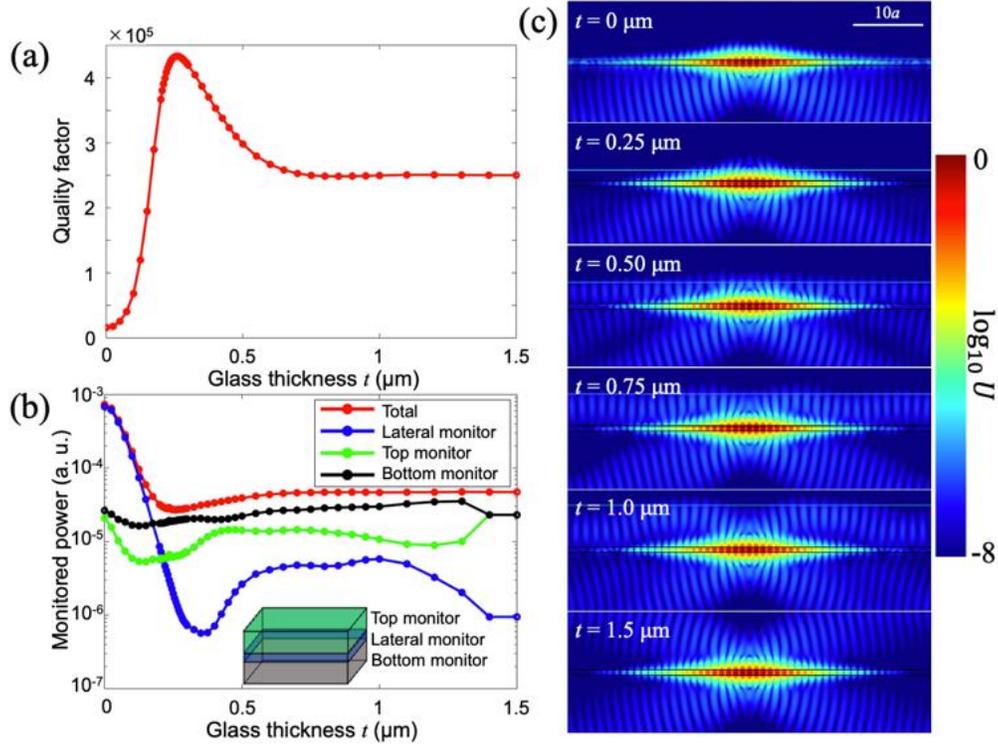

Fig. 2. (a) Simulated $Q$ factors of the hetero-structure nanocavity as a function of glass thickness $t$. (b) Power leakage distributions monitored with the top, side and bottom detectors. The inset shows how the three monitors cover the PhC nanocavity structure. (c) Field distributions of the cavity mode recorded at $x = 0$ calculated for various $t$s. Power density distributions are plotted. The black (blue) lines indicate the position of the PhC slab (the glass-air interface).

The key factor to understand the $Q$ factor improvement at small $t$ is light leakage into the upper direction. The $t$-dependence of power distributions to the top detector (green curve) in Fig. 2(b) suggests weaker light leakage into the upper direction than that into the lower direction. This is due to the presence of the glass thin film, which partially reflects upper radiation from the cavity and converts a part of it to side leakage. This behavior is evidenced in the change of the cross-sectional field distributions when reducing $t$ from 1.5 μm to 0.75 μm. In the latter, the upper radiation appears to be guided in the loaded glass film. For $t$ around 260 nm, one can see an additional dip in the $t$-dependence of the upper power leakage. We interpreted the origin of the dip as the suppression of cavity's radiation due to mode mismatch with optical modes in the upper glass. Indeed, the strong suppression of upper radiation and the absence of lateral waveguiding are observed in the cross-sectional field plot for $t = 250$ nm. The additional suppression of the upper leakage resulted in the peak in the $t$-dependence of $Q$ factor and the increase of $Q$ factor around $t = 260$ nm.

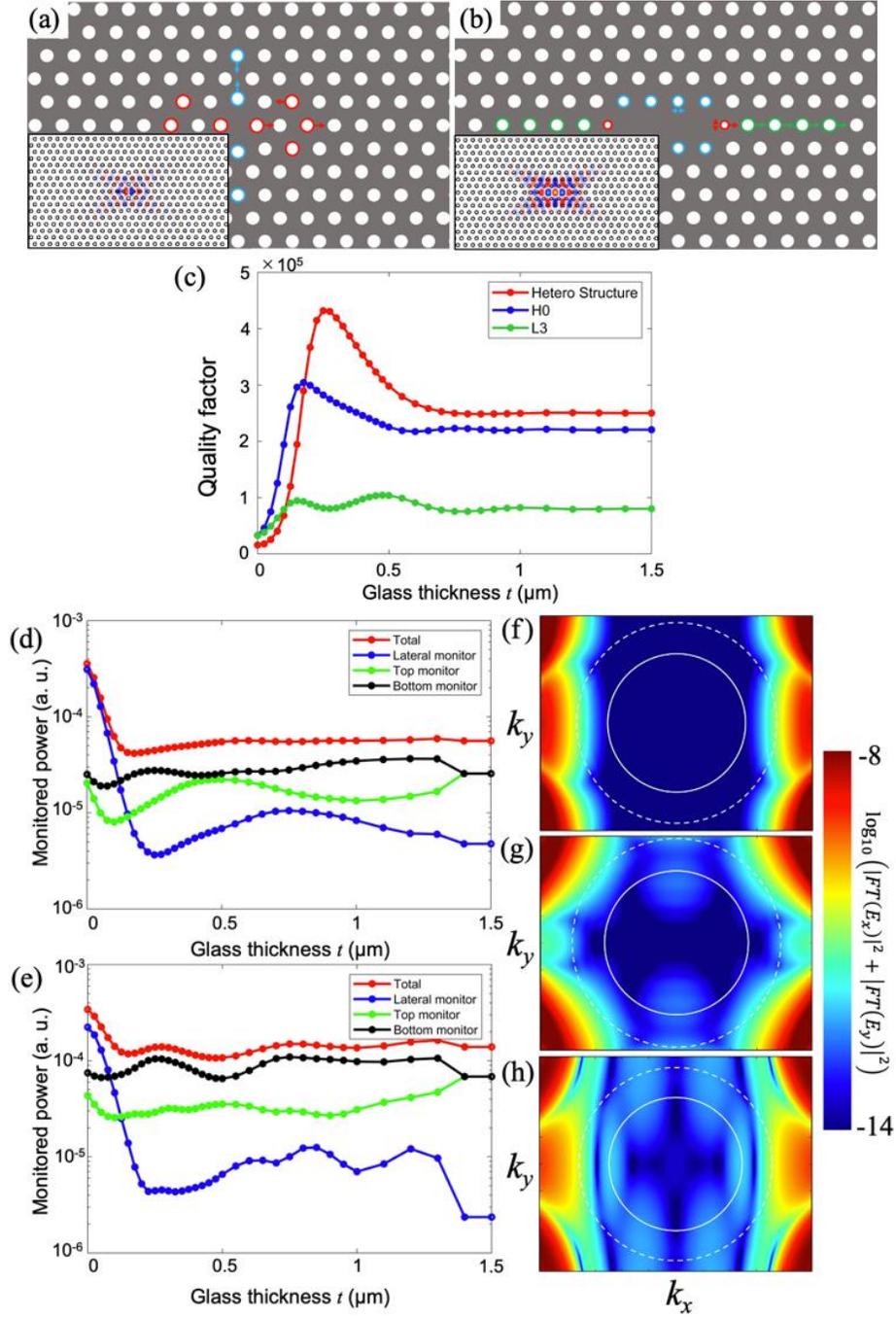

Fig.3. Schematic of the investigated (a) H0 and (b) L3 nanocavity. The positions and radii of the air holes with color are modified mostly in accordance with the references[30,31]. (c) Simulated $Q$ factors of the nanocavities for various glass thicknesses $t$. Computed light leakage distributions of the (d)H0 and (e)L3 nanocavity. Momentum space light intensity distributions of (f)heterostructure, (g)H0 and (h)L3 cavity sandwiched by sufficiently thick glasses. The optical fields in real space were recorded at the slab-glass interface. The color plot shows the distributions of $|FT(E_x)|^2 + |FT(E_y)|^2$, where $FT$ describes the Fourier transform of the real-space field profiles. The solid white lines show the air light lines and the dashed white lines do the glass light lines.

To deepen our understanding of the observed $Q$ factor increase, we also investigated other types of 2D PhC nanocavities covered with glass thin films. Figures 3(a) and (b) show the cavity structures additionally investigated here. The L3 (H0) cavity[30,31] exhibits a high $Q$ factor of $5.6\times10^5$ ($2.7\times10^6$) and a small $V$ of $1.1(\lambda/n)^3$ ($0.34(\lambda/n)^3$) in air. The positions and radii of the air holes in the designs are slightly modified from those in the references. The respective field distributions of the cavity modes are shown in the insets. Figure 3(c) shows a comparison of the $t$-dependences of $Q$ factors for the multi-step heterostructure, H0 and L3 PhC cavities. The H0 cavity shows a similar behavior with the heterostructure cavity: there is a rise of $Q$ factor and its peak value is higher than that in the symmetrically cladded structure. The similarity is also found the power distribution when varying $t$ as shown in Fig. 3(d). The upper leakage curve exhibits a dip for a small $t$ around 200 nm, which results in the peak of $Q$ factor. The suppression of upper leakage and the absence of glass-propagating light are also confirmed in the cross-sectional field plot computed for the H0 (not shown).

In contrast, the $t$-dependence of the L3 cavity shows a largely different behavior. A prominent peak is absent in the curve. Alternatively, multiple weak peaks are observed. The leakage power distribution computed for the L3 cavity in Fig. 3(e) also shows different behavior from those of the heterostructure and H0 cavities. The upper leakage becomes more monotonic, and the lower leakage shows oscillations instead. The $Q$ factor of the L3 cavity is predominantly determined by the lower leakage, which oscillates due to interferences with the light reflected back from the top surface of the upper glass layer. The observed increase of $Q$ factor with respect to the symmetrically cladded structure was only 1.1 times.

To discuss the observed similarity and difference, we computed field distributions of the three cavity modes in the momentum space and plotted as Figs 3(f)-(h). These fields were obtained by Fourier transforming real space fields recorded at the interface of the PhC slab and the glass film. For the heterostucture and H0 cavities, the dominant leaky components lay between the glass and air light lines, which is consistent with the observation that the radiation into modes in the glass largely determines the $Q$ factor of the cavity modes. For the L3 cavity, there are significant components in the air light line, which can be reflected at the surface of the upper glass and interfere with the lower radiation. Meanwhile, there is less field between the glass and air light lines, inhibited the significant increase of $Q$ factor for a thin $t$. The comparison further confirms that the observed increase of $Q$ factor arises from mode mismatch in the upper glass layer for the cavities with oblique radiation with very steep angles.

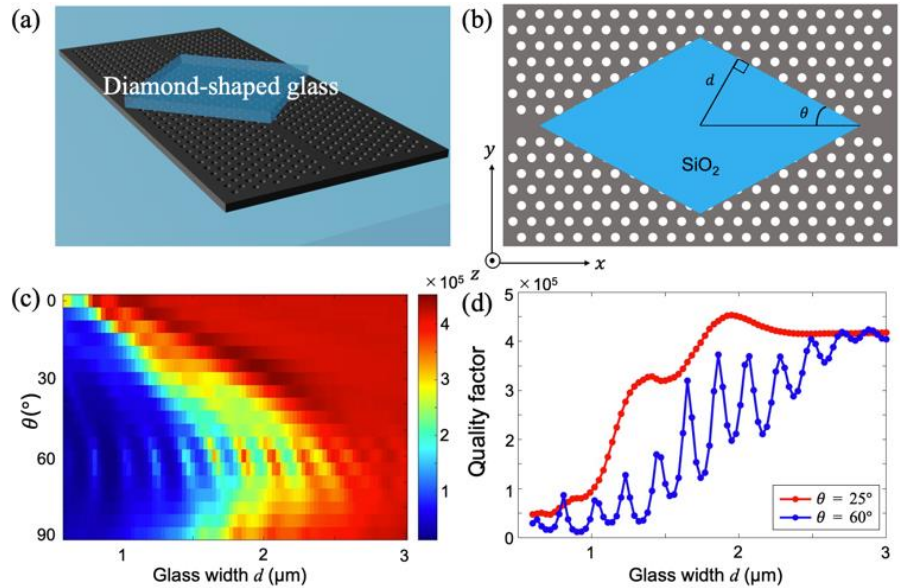

Fig. 4. (a) Bird's-eye and (b) top view of the investigated structure. The upper rhombic glass layer has a fixed thickness of 250 nm, which is close to the optimal value for $d = \infty$. (c) Color plot of $Q$ factors for various $d$ and $\theta$. (d) Computed $Q$ factors when $\theta = 25°$ (red) and $60°$ (blue)

Finally, we study the influence of in-plane structure of the loaded glass film. Figure 4(a) shows a schematic of the structure considered here. Again, we investigated the same multistep heterostructure cavity but the glass film was structured into rhombic. The thickness of the glass was fixed to 250 nm. Figure 4(b) shows the top view of the structure. We varied the glass width $d$ and the rhombic edge angle $\theta$ and calculated $Q$ factor of the cavity mode. Figure 4(c) summarizes the computed $Q$ factors. Overall, $Q$ factor increases for larger $d$ and merges to a saturated value irrespective of $\theta$. We observed oscillatory behaviors of $Q$ factor when increasing d. The clearest oscillation was found for $\theta = 60°$, the line plot of which is shown in Fig. 4(d). The oscillation pitch corresponds to the pitch of the air hole patterned in the GaAs slab. From the analysis of the peak positions, we concluded that the Fresnel reflection of the glass lateral edge becomes stronger when the edge coincides with the air hole. In the case of $\theta = 60°$, we did not find the enhancement of $Q$ factor compared to the saturated value. For smaller $\theta$ around 25°, we observed looser oscillation of $Q$ factor, which can be confirmed in the line plot for $\theta = 25°$ in Fig. 4(d). Interestingly, we observed a 1.1 times enhancement of $Q$ factor when $d = 1.95$ μm. Although the increment is slight, the realization of the high $Q$ factor even by loading such a laterally-small glass is advantageous for high-dense optical integration.

In conclusion, we numerically demonstrated an on-glass 2D PhC nanocavity with a high $Q$ factor of $4.2 \times 10^5$ and a small mode volume of $1.7(\lambda/n)^3$ by covering the structure with a glass thin film. We observed a strong recovery of $Q$ factor by the thin glass due to suppression of the TE-TM mode conversion. We found that the thin-glass layer further increased $Q$ factor due to the suppression of cavity's radiation by mode mismatch, which was confirmed by comparing on-glass heterostructure, H0 and L3 PhC nanocavities topped with thin glass films. Moreover, we clarified in-plane structure of the top glass can additionally increase $Q$ factor. We note that the proposed glass-clad structures can be easily implemented using pick-and-place integration techniques including transfer printing[23,24], and therefore will accelerate the applications of 2D PhC nanocavities in densely-integrated photonics.

**Funding.** KAKENHI ((22H01994, 22H00298, 22H04962), JST FOREST Program (JPMJFR213F).